\documentstyle[preprint,aps]{revtex}
\begin{document}
\title{Causality in dense matter}
\author{B. D. Keister}
\address{
Department of Physics,
Carnegie Mellon University,
Pittsburgh, PA 15213}
\author{W. N. Polyzou}
\address{
Department of Physics and Astronomy,
The University of Iowa,
Iowa City, IA 52242
}
\date{\today}
\maketitle
\begin{abstract}
  The possibility of non-causal signal propagation is examined for
  various theories of dense matter.  This investigation requires a
  discussion of definitions of causality, together with
  interpretations of spacetime position.  Specific examples are used
  to illustrate the satisfaction or violation of causality in
  realistic calculations.
\end{abstract}
\pacs{04.40.Dg,11.90.+t,21.65.+f,97.60.Jd}
\newpage
\narrowtext
\section{Introduction}
An important question for any theory of dense matter is whether that
theory (or an approximate calculation which uses it) permits the
propagation of information faster than the speed of
light~\cite{Shapiro}.  The requirement of Poincar\'e invariance is not
sufficient to guarantee causality, and in any event practical
calculations involve approximations which may also lead to non-causal
properties.  The uncertainty principle leads to additional problems of
interpretation.

The simplest criterion can be obtained by examining sound propagation.
In the long-wavelength limit, small density fluctuations can cause
isentropic pressure-wave propagation whose velocity is given by
\begin{equation}
  v_s^2 = c^2 ({dp\over d\epsilon})_{\vert_{s}},
\end{equation}
where $p$ is the pressure and $\epsilon$ the energy per unit volume.
The wave equation follows from considering energy momentum conservation 
$T^{\mu \nu},_{\nu}=0$ for a linear isentropic perturbation of the 
pressure in a
perfect fluid.  The sound speed follows from the equation of state.
The presence of $c$ is shown explicitly: $p$ and $\epsilon$ may or may
not depend explicitly upon $c$.  Most of the discussion in the
literature concentrates on determining whether specific theories
satisfy $(dp/d\epsilon)_{\vert_{s}} < 1$.  For a system of non-interacting
relativistic fermions, $p={1\over3}\epsilon$.  However, in a
mean-field approximation, Zel'dovich~\cite{Zeldovich} showed that
$\epsilon$ can approach $p$ from below for a system of fermions
interacting via repulsive vector meson exchange.  Bludman and
Ruderman~\cite{BR1,BR2,BR3} also studied questions of causality within
the context of meson-nucleon field theory.  Among other things, they
found that the group velocity can exceed the speed of light in vacuum
($(dp/d\epsilon))_{\vert_s} > 1$), but that real information
propagation does not 
violate causality as long as the index of refraction $n(\omega)$
satisfies the Kramers-Kronig dispersion relation.

It would be useful to gain some insight as to the behavior of dense
systems which employ particle degrees of freedom.  Indeed, dense-matter
calculations which start from a field theory typically end up with
models in which particles interact directly.  In addition, it turns out
that many calculations of the equation of state of nuclear or neutron
matter still satisfy the condition $(dp/d\epsilon)_{\vert_s} < 1$,  even
though they are based on nonrelativistic quantum mechanics.

The primary motivation for this paper is therefore twofold:
\begin{enumerate}
\item Examine causal and non-causal behavior in theories involving
  directly interacting particles.
\item Understand why nonrelativistic calculations work as well as they
  do with respect to causality, in spite of their only approximate
  connection to the principles of special relativity.
\end{enumerate}
To do this, it will be necessary to discuss various definitions of
causality itself, as well as what is meant by spacetime position in a
given theory.  This discussion is presented first, followed by an
examination of specific examples.

\section{Causality and Spacetime Position}

In relativistic quantum mechanics there are two distinct types of
causality.  The first is the requirement that the there is a well posed
initial value  problem.   We call this Cauchy
causality.  The second is the requirement that no signal can propagate
faster than the speed of light.  We call this
Einstein causality.

A violation of Cauchy causality is a serious problem.  In a relativistic
quantum theory the Poincar\'e invariance implies the existence of a
unitary representation of the Poincar\'e group, $U(\Lambda ,a)$ that
acts on the Hilbert space.  The time evolution subgroup 
\begin{equation}
U(t):= U(I, (t,0,0,0))
\end{equation}
relates a state $\vert \psi (t) \rangle$ at time $t$ to an initial state
$\vert \psi (0) \rangle $ at time $t=0$:
\begin{equation}
\vert \psi (t) \rangle = U(t) \vert \psi (0) \rangle.
\label{eq:AA}
\end{equation} 
This shows explicitly that Cauchy causality is a consequence of
Poincar\'e invariance.  The group properties ensure that 
this condition holds in all inertial coordinate systems:
\begin{equation}
U(\Lambda ,a)\vert \psi (t) \rangle = U(\Lambda^{0}{}_{0}t)U(\Lambda ,a)
\vert \psi (0) \rangle 
\end{equation} 
which shows that the transformed final state is also uniquely determined
by time evolving the transformed initial state.  This shows that Cauchy
causality in all inertial coordinates systems is a consequence of
Poincar\'e invariance. Cauchy causality is independent of any
considerations concerning Einstein causality.

This is distinct from the situation in relativistic classical physics,
where Cauchy causality and Poincar\'e covariance imply Einstein
causality.  To see this, assume that events can be precisely identified. 
The Poincar\'e group is defined as the group of transformations that
preserve the proper time between space-time events. The Poincar\'e group
provides an invariant classification of events that are past, future,
and spacelike, relative to a given event. The temporal order of
spacelike separated events is coordinate system dependent. Assume that
the theory satisfies Cauchy causality in all coordinate systems related
by Poincar\'e transformations.  By contradiction, assume that the theory
also violates Einstein causality.  Violating Einstein causality means
that it is possible to send a signal from event $A$ to event $B$ where
$A$ and $B$ are events with a relative spacelike separation.  Since
these events are spacelike separated it is possible to find a coordinate
systems where event $B$ occurs before event $A$.  This leads to a
violation of Cauchy causality in the new coordinate system,
contradicting the assumption that the theory violates Einstein
causality.  Thus, we conclude that if we can precisely measure all
spacetime events and the theory satisfies Cauchy causality in all
coordinate systems related by Poincar\'e transformation, then the theory
must also satisfy Einstein causality.

In a classical {\it dynamical} theory, the situation is not as clear
as suggested by the previous paragraph.  For instance, it is known
that is impossible to construct an interacting canonical Hamiltonian
theory with a non-trivial representation of the Poincar\'e group (with
the Lie algebra realized in terms of classical Poisson brackets) that
has spacetime events that transform covariantly~\cite{Currie}.  This
gives some clues concerning the difficulties that occur in the quantum
case.

In the quantum case, the Poincar\'e symmetry is
preserved, but the concept of an event that occurs at a fixed
spacetime point
cannot be precisely formulated in terms of particle degrees of freedom.
Specifically, one cannot use the complete set of commuting operators
that define the state of a single particle to construct a set of
commuting  Hermitian operators whose eigenvalues transform like the
spacetime coordinates of an event.  More specifically, the uncertainty
principle in relativistic quantum mechanics does not permit one to
localize a particle with arbitrary precision.  Its position can only be
determined up to an uncertainty on the scale of the particle's Compton 
wavelength.   A local field operator is  needed for a precise
formulatation of
Einstein causality.
   
All of these issues can be easily illustrated for the case of a free
spinless quantum particle of mass $m$.   For a free particle, the
momentum eigenstates
\begin{equation}
\vert {\bf p}: \, \rangle \qquad \langle {\bf p} \, \vert {\bf p}\,'
\rangle =  (2\pi)^3
\delta^3 ({\bf p} - {\bf p}\,' )  
\label{eq:AB}
\end{equation}
form a complete set of states. The unitary representation of the
Poincar\'e group, 
\begin{equation}
U(\Lambda ,a) \vert {\bf p}\, \rangle = \sqrt{{\omega_m ({\bf p}') \over
\omega_m ({\bf p}) }} e^{- i \Lambda p \cdot a} \vert {\bf p}\,' \rangle, 
\label{eq:AC}
\end{equation}
is determined up to an overall phase by unitarity, the mass and spin of
the particle,  and the requirement that $p:=(\omega_m ({\bf p} \,) ,
{\bf p}\,)$ transforms like a four vector.
 
This representation of the Poincar\'e group contains the time evolution
subgroup corresponding to a free particle of mass $m$.  It leads to a
well posed initial-value problem that is consistent in any frame of
reference.  The single-particle dynamics thus satisfies Cauchy causality
and Poincar\'e invariance.

Classically, an event $({\bf x},t)$ transforms like a four-vector under
Poincar\'e transformations.  In order to investigate Einstein causality,
assume by contradiction that it is possible to find a single particle
state $\vert {\bf x},t \rangle $ corresponding to the particle being
found at the point ${\bf x}$ at a given time $t$. The requirement that
$\vert {\bf x},t \rangle$ transform like a classical event is 
\begin{equation}
U(\Lambda ,a) \vert {\bf x} ,t \rangle = \vert {\bf x}\,', t' \rangle ,
\label{eq:AD}
\end{equation}
where 
\begin{equation}
({\bf x}' ,t') := \Lambda ({\bf x} ,t ) + a .
\label{eq:AE}
\end{equation}
It follows from (\ref{eq:AC}) and (\ref{eq:AD}) using spacetime
translations
that  
\begin{equation}
\langle {\bf p}\, \vert {\bf x} ,t \rangle =
\langle {\bf p}\, \vert U(I, ({\bf x},t) \vert {\bf 0} ,0 \rangle =
e^{i {\bf p} \cdot {\bf x} - i \omega_m ({\bf p}\, ) t} 
\langle {\bf p}\, \vert {\bf 0} ,0 \rangle 
\label{eq:AF}
\end{equation}
and using Lorentz transformation that 
\begin{equation}
\langle {\bf p}\, \vert {\bf 0} ,0 \rangle =
\langle {\bf p}\, \vert U(B(p) ) \vert {\bf 0} ,0 \rangle = 
\sqrt{{m \over \omega_m ({\bf p} \,)} }
\langle {\bf p}={\bf 0} \vert {\bf x}={\bf 0},t=0 \rangle =
N \sqrt{{m \over \omega_m ({\bf p} \,)} }
\label{eq:AG}
\end{equation}
where $N$ is a normalization constant.   Combining (\ref{eq:AF}) and 
(\ref{eq:AG}) implies the state $\langle {\bf p}\, \vert {\bf x}\rangle$
necessarily has the form of a plane wave state 
\begin{equation}
\langle {\bf p}\, \vert {\bf x} ,t \rangle =
N \sqrt{{m \over \omega_m ({\bf p} \,)}}
e^{i {\bf p} \cdot {\bf x} - i \omega_m ({\bf p}\, ) t} .
\label{eq:AH}
\end{equation}  
To check Einstein causality we calculate the overlap of states
corresponding to different space-time points:
\begin{eqnarray}
  \label{eq:AHA}
  \langle {\bf x} , t \vert {\bf x}\,' ,t' \rangle &&=
  \vert N \vert^2 {1\over(2\pi)^3}
  \int {d{\bf p} \over 2\omega_m ({\bf p}\, ) }
  e^{ i {\bf p} \cdot ({\bf x}\,' - {\bf x}\,) - 
    \omega_m ({\bf p}\,' ) (t' -t) } \nonumber \\
  &&=  -i  \vert N \vert^2 D^-(x'-x),
\end{eqnarray}
where $D^-(x)$ is the negative frequency part of the Pauli-Jordan
commutator function, which has the form~\cite{Bogoliubov}
\begin{eqnarray}
  \label{eq:AI}
  D^- (x) &&:= {i\over (2\pi)^3} \int {d{\bf k}\over 2\omega_m({\bf k})}
    e^{-ik\cdot(x-y)} \nonumber \\
  &&= {1 \over 4 \pi} \epsilon (x^0) \delta ( \tau ) 
  - {m \theta (\tau^2 )\over 8 \pi \tau} [ 
  \epsilon (x^0) J_1 (m
  \tau) -i N_1 (m\tau ) ] + 
  {m \theta (-\tau^2 ) \over 4 \pi^2 \sqrt{-\tau^2}} \theta  K_1 (m
  \sqrt{-\tau^2}) ,
\end{eqnarray}
and  $\tau$ is the proper time
between the events.   This does not vanish for spacelike separation
($\tau^2 < 0$).  The Bessel function $K_1 (x)$ has the asymptotic form
\begin{equation}
K_1 (x) \to \sqrt{{\pi \over 2x}}e^{-x}(1 + {3 \over 8x} + \cdots ),
\end{equation}
which decays exponentially for
large $x$.  This suggests an apparent violation of Einstein causality 
over a range given by the Compton wavelength of the particle.  Note that
what appears in the exponent is $-m\sqrt{-\tau^2}$, which involves the
invariant spacelike separation.  The actual spatial separation can be
much larger than the Compton wavelength, provided the time difference 
between the events is also large.  The problem is not with Einstein
causality; it is related to the non-existence of a position operator
that transforms as the space components of a four vector, or, equivalently,
to the non-existence of a state vector representing a point particle. 

The problem is that if one interprets the Fourier transform of the
momentum space wave function as a position probability amplitude, then the
position operator ${\bf x}= i{\bf \nabla}_p$ is 
related to the generator of Lorentz boosts by equation
(\ref{eq:AC}), which also must be expressible in terms of 
the derivatives with respect to the momentum.  For spinless
particles, this relation is    
\begin{equation}
{\bf x} := -{1 \over 2} \lbrace {1 \over H}, {\bf K} \rbrace, 
\label{eq:AJ}
\end{equation}
with slightly more complicated expressions for particles with spin.   The
transformation properties of the Hamiltonian $H$ (zeroeth component of a
four vector) and the boost generator ${\bf K}$ ($0i$ components of a
rank two antisymmetric tensor) determine the transformation properties
of ${\bf x}$ under Lorentz transformation.  

To see that the problems with defining a position operator have 
nothing to do with Einstein causality, consider
the free local scalar field $\phi (x)$.  This operator satisfies the
local commutation relations 
\begin{equation}
[\phi (x) , \phi (y) ]_-=[\phi^{\dagger} (x) , 
\phi^{\dagger} (y) ]_-=[\phi^{\dagger} (x) , \phi (y) ]_-=0
\label{eq:AK}
\end{equation}
when $x-y$ is spacelike, and the covariance conditions
\begin{equation}
U(\Lambda ,a ) \phi (x) U^{\dagger} (\Lambda ,a) = \phi (\Lambda ,x +a);
\label{eq:AL}
\end{equation}
\begin{equation}
U(\Lambda ,a ) \phi^{\dagger} (x) U^{\dagger} (\Lambda ,a) = \phi^{\dagger}
(\Lambda ,x +a).
\label{eq:AM}
\end{equation}
The field operator can be used to construct observables 
\begin{equation}
\phi (f) = \int d^4 x f(x) \phi (x),
\label{eq:AN}
\end{equation}
corresponding to a space-time region ${\cal O}$ where $f(x) =0$ for $x
\notin {\cal O}$.  The underlying quantum theory satisfies Einstein
causality (\ref{eq:AK}).  
However, if we define the state $\vert x \rangle := \phi (x)
\vert 0 \rangle$ it follows from (\ref{eq:AL}) that $\vert x \rangle$
satisfies (\ref{eq:AD}).  By direct computation, 
\begin{equation}
\langle x \vert y \rangle = -i D^- (x-y) ,
\end{equation}
where $\tau^2 = -(x-y)^2$.
This expression is identical (up to a constant multiplicative factor) to
(\ref{eq:AHA}).  In this case the result is derived in local field
theory, which clearly satisfies Einstein causality.  The key
observation is that $\vert x \rangle$ is a single particle state, but
the argument $x$ does not correspond to a precise space-time position of
the particle.

What is relevant in the case of the field is that antiparticle degrees
of freedom are needed ensure the commutation relations (\ref{eq:AK}). 
The point is that the relevant local observables are the fields -  not
the particle excitation of the field.  Although it is possible to
localize the field, the single particle eigenstates are not localizable,
even for the structureless particles of a free field theory.  What
distinguishes a particle theory from a field theory is that it is a
theory of an infinite number of degrees of freedom.   Particle  theories
do not contain enough operators to separate points in arbitrarily small
spacetime regions.

Thus, we see that in a theory of particles or fields, the particle
degrees of freedom cannot be used as a reliable test of Einstein causality
when the separation between particles is on the order of a Compton
wavelength of the particle.  Nevertheless,  it is possible to establish
a violation of Einstein causality if it occurs on a scale significantly
larger than a Compton wavelength of the particle.  Certainly spacetime
positions of particles are determined experimentally to within
uncertainties, although large compared to the particles Compton
wavelength, that are sufficiently small to localize the particle within
experimental uncertaintites. 

If we start with a localized disturbance, and evolve it in time with a
non-local Hamiltonian,  then we can test to see the extent to which the
disturbance propagates in a manner that violates Einstein causality.  
Specifically, if non-causal effects exist on scales beyond the
uncertainties in position, then the theory can ultimately lead to non-causal
effects. 

We now illustrate how the addition of an interaction affects Einstein
causality. 
The simplest illustration is to consider the first order effect due to
the interaction in a Bakamjian-Thomas~\cite{BT} two-body
model~\cite{KP}.   The effect of 
the interaction can be treated using perturbation theory.  Let $H=H_0+V$,
where $V$ is the difference between the full Bakamjian-Thomas
Hamiltonian and the kinematic Hamiltonian.  The time evolution operator
can be expanded in powers of the interaction using the formula
\begin{equation}
e^{-iHt} = e^{-iH_0 t} + \sum_{n=1}^\infty S_n (t),
\label{eq:AP}
\end{equation}
where
\begin{equation}
S_0 (t) = e^{-iH_0t} \qquad S_k (t) = -i \int_0^t e^{-i H_0
(t-s)}VS_{k-1}(s) ds.
\label{eq:AQ}
\end{equation}
The first order correction is
\begin{equation}
S_1 (t) = -i \int_0^t e^{-i H_0 (t-s)}Ve^{-iH_0 s} ds.
\label{eq:AR}
\end{equation}
If we let ${\bf x}_1$ and ${\bf x}_2$ be the Fourier transforms of the 
single particle momenta and let 
\begin{equation}
e^{iH_0t} = e^{i(H_1+H_2)t} = U_1 (t) U_2 (t) ,
\label{eq:AS}
\end{equation}
then we can write
\begin{eqnarray}
  \label{eq:AT}
  \langle {\bf x}_1 , {\bf x}_2 \vert S_1 (t) \vert {\bf x}_1\,',
  {\bf x}_2\,' \rangle &&=
  -i \int_0^t ds \int \langle {\bf x}_1 \vert U_1 (t-s) \vert
  {\bf x}_1\,'' \rangle   
  \langle {\bf x}_2 \vert U_2 (t-s) \vert {\bf x}_2\,'' \rangle  
  d{\bf x}_1'' d{\bf x}_2'' \nonumber \\
  &&\qquad\times
  \langle {\bf x}_1\,'' , {\bf x}_2\,'' \vert V
  \vert {\bf x}_1\,'' , {\bf x}_2\,''\rangle
  d{\bf x}_1''' d{\bf x}_2''' \nonumber \\
  &&\qquad\times
  \langle {\bf x}_1\,''' \vert U_1 (t-s) \vert {\bf x}_1\,' \rangle  
  \langle {\bf x}_2\,''' \vert U_2 (t-s) \vert {\bf x}_2\,' \rangle  .
\end{eqnarray}
In this form, it is easy to examine the causality properties. The
operators $U_i(t)$ are free-particle time evolution operators.   The
coordinates ${\bf x}_i$ describe the position of the particle up to an
uncertainty that is about the size of the Compton wavelength of the
particle.  The matrix element  $\langle {\bf x}_i \vert U_i (t) \vert
{\bf x}_i' \rangle$ can be computed:
\begin{equation}
\langle {\bf x} \vert U_1 (t-t') \vert {\bf x}\,' \rangle =
2 {\partial \over \partial t'}D^- (x'-x),
\end{equation}
which again vanishes exponentially when $\vert {\bf x}_i - {\bf x}_i\,'
\vert^2 - t^2 $ is much larger than the square of a Compton wavelength. 
The interaction $V$ in (\ref{eq:AT}) is instantaneous. Thus propagation
from ${\bf x}_i\,'$ to ${\bf x}_i'''$ in a time $s$ is causal up to
uncertainties in position.  The propagation  from ${\bf x}_i\,'''$ to
${\bf x}_i\,''$ is instantaneous over the range of the interaction, and
finally the propagation from ${\bf x}_i\,''$ to ${\bf x}_i$ in time
$t-s$ is causal up to uncertainties in position.  It follows that  the
range of the interaction sets the scale on which non-causal effects
occur in perturbation theory.  If the range of the interaction  is much
larger than the Compton wavelength of one of the particles, then it
should in principle be possible to detect such non-causal effects
experimentally.  This analysis is based on perturbation theory and only
gives an indication of what can happen.  Exact calculations can be
easily performed.

Many of these issues arise in a many-body system.  Questions of testing
causality are intertwined with questions related to the interpretation of
position variables.   These issues are more complicated if the
particles are detected in the nuclear medium.

Although it is worthwhile to understand these issues in a many-body
problem,  for the purpose of testing for violations of Einstein
causality in particle theories, it is sensible to utilize an idealized
model where all of the questions of interpretation become irrelevant. 
Consider a dilute gas of particles and imagine a long chain of
successive scattering events where the mean separation of the particles
is much larger than either the range of the interaction or the Compton
wavelength of the particles.  If each successive interaction leads to a
small violation of causality, then it follows that this violation can be
amplified to any desired amount by a sufficiently long chain of such
scattering events.    In this picture, many-body effects are unimportant.
The amplification allows one to obtain violations of causality that
occur on macroscopic scales.  For a sufficiently dilute gas, we
detect free particles with no medium effects.  The spacetime position of
the detected particle can be determine up to experimental uncertainty. 
With sufficient amplification, the causality violations can be made large
on a scale that is large compared to experimental uncertainties.  

In this setting, Einstein causality can be reduced to
the study of the two-body problem, which can be dealt with in a
practical manner.

This type of study can be used to establish the possibility of being
able to establish violations of Einstein causality in particle models.
In dense matter, there are addidtional effects that require 
additional attention.  Most notably among are the effects of the Pauli
principle and cluster properties.  The Pauli principle restricts the 
allowable final states in a scattering reaction, while cluster
properties requires the existance of a number of many-body interactions. 
The role of these effects is difficult to estimate without a specific
model.

\section{Long-Wavelength Limit}
As noted in the Introduction, the velocity of so-called zero-sound
propagation is governed by the pressure-energy relation $(dp/d\epsilon)_s$,
where $p(\epsilon)$ is the equation of state.  (We assume for this
discussion that the system has zero temperature.)  The nonrelativistic
expression,
\begin{equation}
  \label{NRvelocity}
  v^2 = (dp/d\rho)_s,
\end{equation}
where $\rho$ is the mass density,
differs from the relativistic expression by the replacement
$\rho\to\epsilon/c^2$. 

The usual procedure is to obtain a relation between the energy density
$\epsilon$ and the number density $n$ in terms of a dynamical theory.
The pressure is then related to the energy density
$\epsilon$ and the number density $n$ via
\begin{equation}
  \label{evsn}
  p = n {d\epsilon\over d n} - \epsilon.
\end{equation}
A test of causality is to see whether the resulting equation of state
$p(\epsilon)$ satisfies $dp/d \epsilon < 1$ (at least for physically
reasonable values of $\epsilon$).

Nonrelativistic calculations of the nuclear equation of state have very
minimal dependence upon the velocity of light.  The energy density can
be written as
\begin{equation}
  \label{NRepsilon}
  \epsilon = nmc^2 + \Delta\epsilon_{NR},
\end{equation}
where $m$ is the mass of the constituent particle and $\Delta\epsilon_{NR}$
includes the kinetic and interaction energy 
resulting from a calculation which does not
depend explicitly upon $c$.  From Eq.~\ref{evsn}, it can be seen that
the rest energy $nmc^2$ does not contribute to the pressure, and thus
the pressure does not depend upon $c$.  The quantity ${dp\over d\epsilon}$  
does depend upon $c$ via the energy density:
\begin{equation}
  \label{eq:NRdpde}
  {dp\over d\epsilon} = \left[ nmc^2
    + n{d\Delta\epsilon_{NR}\over dn}\right]^{-1}
    n^2 {d^2\Delta\epsilon_{NR}\over dn^2}.
\end{equation}
The appearance of $c$ in Eq.~\ref{eq:NRdpde} is trivial unless
$d\epsilon_{NR}/dn$ is comparable to $mc^2$, but in that case a
nonrelativistic approximation is no longer valid.  Thus, the test
whether $dp/d\epsilon < 1$ may be less severe than examining the
internal consistency of the kinematics leading to the equation of
state for nonrelativistic theories of dense matter

\section{Beyond the Long-Wavelength Limit}
   
\subsection{Particle theories and wave packet propagation}

Local quantum field theories are not necessary for describing an
interacting system of particles consistent with Poincar\'e invariance and
quantum mechanics.  One can also formulate models using a Hamiltonian,
in which particles are the designated degrees of freedom.  The dynamics
can be specified in terms of a series of direct interactions among
$2,3,\dots$ particles.  Cluster properties requires that widely
separated clusters of particles behave as they would in complete
isolation, the presence of momentum dependent many-body interactions is
a consequence.  

A strongly interacting relativistic many-body system at high density 
would therefore satisfy Poincar\'e invariance and macroscopic causality,
but in general it would not satisfy the microscopic causality that is 
characterized by local interacting fields.  Nevertheless, the physical test
of such a system is whether it permits signal propagation faster than
the velocity of light.  Bludman and Ruderman suggest that this can be
tested by examining the poles of the Green function of the interacting
system (the sound modes). However, such a Green function is difficult to
calculate for a many-body system.

Our purpose in this study is to understand the conditions where a
breakdown in causality (in the sense of signal propagation) might
occur.  To get a preliminary picture, we consider a system of
interacting particles in one dimension.  One can imagine an incoming
projectile striking a constituent particle, which in turn recoils
forward, striking a second constituent, which in turn strikes a third,
{\it etc.} Classically, if the particles interact via a rigid barrier
({\it e.g.,} a string of croquet balls), then causality is violated in
the sense that each particle responds instantly to the near approach of
a projectile, that is, the time of light propagation across the barrier
is not taken into account.  For a quantum mechanical system,
information propagates via a wave packet.  

We will consider two properties of causality with respect to wave
packets: the spacetime structure of packet propagation, and the
overall time advance of a packet.

In order to study the spacetime structure of packet propagation, it is
useful to examine the negative-frequency portion of Pauli-Jordan
function in one dimension: 
\begin{eqnarray}
  \label{PauliJordan}
  D_{1}^{-}(x) &&= {i\over2\pi} \int d^2k\, e^{-ik\cdot x} \delta(k^2 - m^2)
  \theta(k^0) \nonumber \\
  &&= {1\over 2\pi i} \int_{-\infty}^{+\infty} dk \, {1\over2\omega_m(k)} 
  e^{ikx} e^{-i\omega t}.
\end{eqnarray}
After a variable change,
\begin{equation}
  \label{k_hyper}
  k = m \sinh u; \quad \omega = m \cosh u; \quad dk = \omega du,
\end{equation}
we have
\begin{equation}
  \label{PauliJordanHyper}
  D_{1}^{-}(x) = {i\over2\pi}{1\over2} \int_{-\infty}^{+\infty} du \,
  e^{i m (x\sinh u - t \cosh u)}.
\end{equation}
For the case $x>0, t>0, t>x$ (timelike interval), the integral can be
reduced as follows:
\begin{equation}
  \label{x_hyper}
  x = \tau \sinh z; t = \tau \cosh z,
\end{equation}
\begin{equation}
  \label{PauliJordantimelike}
  D_{1}^{-}(x) = {i\over2\pi}{1\over2} \int_{-\infty}^{+\infty} du \,
  e^{-i m \tau\cosh u} = {1\over4} H_0^{(2)}(m\tau),
\end{equation}
where $H_0^{(2)}$ is the zeroeth-order Hankel function of the second
kind.  
For the case $x>0, t>0, t<x$ (spacelike interval), the integral can be
reduced as follows:
\begin{equation}
  \label{t_hyper}
  x = \lambda \cosh z; t = \lambda \sinh z,
\end{equation}
\begin{equation}
  \label{PauliJordanspacelike}
  D_{1}^{-}(x) = {i\over2\pi}{1\over2} \int_{-\infty}^{+\infty} du \,
  e^{i m \lambda\sinh u} = {i\over2\pi} K_0(m\lambda),
\end{equation}
where $K_0$ is the zeroeth-order modified Bessel function.  For
spacelike separation, $D^{-}$ does not vanish identically, but rather
falls off with a range given by the particle Compton wavelength.

In the examples which follow, we will encounter integrals similar to
that for $D^{-}$, of the form
\begin{equation}
  \label{PauliJordanmodified}
  \int_{-\infty}^{+\infty} dk \,
  f(k) e^{ikx} e^{-i\omega t}.
\end{equation}
Even if the integral cannot be evaluated analytically, we can gain some
insight by considering the integrand in the complex plane.  For
spacelike separation ($x>0, t>0, x>t$), the phase factor vanishes
exponentially as $|k|\to\infty$ in the upper half plane.  If $f(k)$
is analytic in the upper half plane, then the
integration contour can be distorted away from the real axis, with the
contour ``pinned'' around the branch cut beginning at $k=+im$.  The
falloff of the result will therefore be limited by the factor
$e^{-mx}$, {\it i.e.,} the Compton wavelength scale.  This qualitative
argument applies to functions $f(k)$ which do not have additional
exponential behavior in $k$.  If $f(k)$ has poles or branch points in
the upper half plane, then there may be additional scales determining
the rate of falloff.  In any event, the rate of falloff will be no
sharper than the Compton wavelength.

The time evolution of a wave packet is given by
\begin{equation}
  \label{time_evolution}
  \Psi(x',t') = e^{-iHt'} \Psi(x,t=0).
\end{equation}
This expression can be evaluated by inserting a complete set of states.
In the center of momentum, assuming no bound states, the sum is saturated
by scattering states $| k^+ \rangle$:
\begin{equation}
  \label{completeness}
  \Psi(x',t') = 
  \int_{-\infty}^{+\infty} {dk\over2\pi}\, \langle x' | k^+ \rangle 
  e^{-2i\sqrt{m^2 + k^2} \,t'} 
  \langle k^+ | x \rangle \Psi(x,t=0).
\end{equation}
The key element in the structure of wave packet propagation is the
spectral integral in Eq.~\ref{completeness}:
\begin{equation}
  \label{spectralsum}
  \int_{-\infty}^{+\infty} {dk\over2\pi}\, \langle x' | k^+ \rangle 
  e^{-2i\sqrt{m^2 + k^2} \,t'} 
  \langle k^+ | x \rangle.
\end{equation}
For non-interacting particles, it has the form of the modified
Pauli-Jordan function, as defined in Eq.~\ref{PauliJordanmodified}.
For interacting systems, we will collect exponential factors in order
to use the analysis following Eq.~\ref{PauliJordanmodified}.

We can learn about information propagation through a dense system by
studying the scattering of two of its constituents.  The relevant
question for the series of packet backscatterings described above is
whether the backscattered packet arrives at a given spatial point
``too early'' in terms of light propagation time across the range of a
direct interaction.  This {\it time advance} of the backscattered
packet is in principle an observable effect: each pairwise scattering
contributes a time advance $\Delta T_{adv}$, which can accumulate over
many scatterings.  Whether one can observe a violation of causality on
a macroscopic scale depends upon whether information travel time $T$
over a macroscopic distance $L$ is less than $L/c$ when $T = L/v -
N\Delta T_{adv}$, where $N$ is the number of pairwise scatterings.
Observable consequences would require both $v$ close to $c$ and
$Nv\Delta T_{adv}$ close to $L$ (a dense system).

The time advance $\Delta T_{adv}$ is determined from the Wigner
time delay~\cite{Wigner,Newton} $\Delta T_{delay} = - \Delta T_{adv}$,
which can be 
expressed in terms of the $S$ matrix~\cite{JauchMarchand} via
\begin{equation}
  \label{delay}
  \Delta T_{delay} = -i{1\over v}
  \sum_{\pm}  \langle \pm | S(k) | + \rangle^*
  {d\over dk} \langle \pm | S(k) | + \rangle,
\end{equation}
where $\langle \pm | S(k) | + \rangle$ is the $S$ matrix for
scattering of a particle incident from $x\to+\infty$ to
$x\to\pm\infty$.

\subsection{Examples}
\subsubsection{Potential Barrier}
A simple example is the potential barrier:
\begin{equation}
  V(x) = 
  \left\{ \begin{array}{ll}
      V_0, & -a \le x \le 0 \\
      0, & x < -a,\quad x > 0 \\
    \end{array}  \right.
  \label{barrier}
\end{equation}
The solution is
\begin{equation}
  \label{barriersolution}
  \langle x | k^+ \rangle = \psi_k^+(x) = 
  \left\{
  \begin{array}{ll}
    e^{ikx} + A e^{-ikx},\qquad & x < -a \\
    B e^{ilx} + C e^{-ilx},\qquad & -a \le x \le 0 \\
    D e^{ikx},\qquad & x > 0 \\
  \end{array}
  \right.
\end{equation}
where
\begin{eqnarray}
  \label{ABCD}
  B &&= {2(1+\nu)e^{-ika}\over\Delta}; \qquad
  C = -B\left({1-\nu\over 1+\nu}\right); \nonumber \\
  A &&= {\textstyle{1\over2}}e^{-ika} B(1-\nu)
  \left(e^{-ila} - e^{ila}\right); \qquad
  D = B + C; \nonumber \\
  l^2 &&= k^2 - mV_0; \qquad \nu = l/k; \qquad
  \Delta = (1+\nu)^2 e^{-ila} - (1-\nu)^2e^{ila}.
\end{eqnarray}
For the case $k^2 < V_0$, $l\to i\lambda$, where $\lambda^2 = mV_0 -
k^2$. 

The simplest case to consider is the rigid barrier: $V_0\to\infty$.
The solution can be written as
\begin{equation}
  \label{rigid}
  \psi_k^+(p) = A \sin k(x+a) \theta(x+a).
\end{equation}
In the spectral sum of Eq.~\ref{spectralsum}, there will appear
exponentials associated with free particles,
\begin{equation}
  e^{i[\pm k(x - x') - 2\sqrt{m^2 + k^2} \,t']},
\end{equation}
as well as exponentials associated with the barrier:
\begin{equation}
  e^{i[\pm k(x + x' + 2a) - 2\sqrt{m^2 + k^2} \,t']}.
\end{equation}
The spectral sum over this second set of exponentials will yield a
decaying function of range $1/m$ under the condition
\begin{equation}
  (x + x' + 2a)^2 > 4t'{}^2.
\end{equation}
Thus, there is a range of spacetime events where $(x + x' + 2a)^2 <
4t'{}^2$, but also $(x + x')^2 > 4t'{}^2$ (the actual physical causality
condition), for which the spectral sum yields oscillating rather than a
decaying behavior.  This range corresponds precisely to the region of
overlap of the barriers.  Physically, two hard spheres will repel each
other at the instant their edges come in contact.  The struck sphere
starts moving instantly, leaving no time for information about the
interaction to travel across the sphere at the speed of light.

The non-causal behavior illustrated above has a range characterized by
the interaction, and goes beyond the Compton wavelength associated with
particle masses.  How does this non-causal behavior manifest itself in
terms of information propagation in a dense system?  One way to see
this is to examine the propagation of a wave packet which represents a
pulse of information.  Consider the evolution of a Gaussian packet:
\begin{equation}
  \label{Gaussian}
  \Psi(x, t=0) = {1\over{\sqrt{\sqrt{2\pi}b}}} e^{-x^2 / 2 b^2}.
\end{equation}
The prototype for successive scatterings through a medium is the
backscattered portion of a packet in a two-particle subsystem in its
center of momentum, as shown in Fig.~1.  For a causal system, we expect
the centroid of a backscattered packet to arrive at a point $x'$ at a
time $t' = (-x'-x)/v$, where $v = k/2\omega_k$ is the velocity in the
center-of-momentum frame.  Note that both $x$ and $x'$ are negative.  The
time advance given by Eq.~\ref{delay} is $\Delta T_{adv} = 2a/v$,
{\it i.e.,} the travel time for the packet to go from $x=-a$ to $x=0$ and
back again, which is eliminated by the presence of the rigid barrier.
We have evaluated numerically the packet propagation for various
parameters $k$, $a$, and $b$, and find that in all cases considered,
backscattered packets reform almost immediately upon collision with
the barrier, and reflect the time advance $2a/v$.

For the case of a {\it finite} barrier, the coefficients $A, B, C, D$
have a complicated dependence on $k$ through $l$ or $\lambda$, so it
is more difficult to formulate a simple analytic picture.  Numerical
evaluation of Eq.~\ref{delay} yields a time advance, though the amount
is smaller than $2a/v$.  

\subsubsection{Separable Yamaguchi Potential}
The scattering wave function $\psi_k^+(p) = \langle p | k^+ \rangle$
satisfies the equation
\begin{equation}
  \psi_k^+(p) = \phi_k(p) 
  + \int_{-\infty}^{+\infty} {dq\over2\pi}\,  G_+(p) V(p, q) \psi_k^+(q),
\end{equation}
where the Green function reflects the choice of adding interactions to
the non-interacting mass squared:
\begin{equation}
  G_+(p) = {1\over 4(m^2 + k^2) - 4(m^2 + p^2) + i0}
       = {1\over 4(k^2 - p^2) + i0}
\end{equation}
and
\begin{equation}
  V(p, q) = g v(p) v(q)
\end{equation}
has a separable Yamaguchi form.  For definiteness, we use
\begin{equation}
  \label{pole}
  v(p) = {1\over p^2 + \Lambda^2}.
\end{equation}
Using the Yamaguchi pole form, we get
\begin{equation}
  \label{xwave}
  \langle x | k^+ \rangle = e^{ikx}
  - {g v^2(k) \over D(k)} {i\over 8}
  \left[{1\over k} e^{ik|x|} + {i\over \Lambda} e^{-\Lambda |x|} \right],
\end{equation}
where
\begin{equation}
  D(k) = 1 + {ig\over 4} v^2(k)
  \left[{1\over 2k} + {i\over 4 \Lambda^3} (k^2 + 3\Lambda^2)\right].
\end{equation}
Equation~\ref{xwave} is the sum of the incident wave, plus a scattered
wave containing two pieces, one obeying ``outgoing'' boundary
conditions (in one dimension), the other having an exponential falloff
outside the interaction region.

We now consider the spectral sum in Eq.~\ref{spectralsum}.  All of the
$x$ dependence appears as explicit exponentials in Eq.~\ref{xwave}.
The oscillating exponentials, together with the time dependence,
yield exponentials of the form
\begin{equation}
e^{i[k(x - x') - 2\sqrt{m^2 + k^2} \,t']}.
\end{equation}
While there is additional
$k$ dependence multiplying this overall exponential, it has the form
given by Eq.~\ref{PauliJordanmodified}, and we thus expect the usual
causality condition associated 
with the $D$ and $K$ functions~\cite{Bogoliubov} to arise.
However, there will also appear cross terms involving oscillating
exponentials and decaying exponentials, as well as a bilinear term of
the form
\begin{equation}
  e^{-\Lambda x} e^{-\Lambda x'}.
\end{equation}
The integration variable $k$ does not
appear in these exponentials, but they do affect the causality
condition.  For sufficiently small values of $t$, therefore, these
terms will contribute a non-vanishing amount in the acausal region.
This acausal effect vanishes for values $x$ and/or $x'$ greater than
the scale $\Lambda^{-1}$.

While one can see explicitly how this works for a separable Yamaguchi
interaction, it seems clear that the real issue is one of scales.
Similar effects should occur in more realistic three dimensional
quantum mechanical models.

How will a wave packet be affected?  While the packet overlaps the
interaction region, there will certainly be non-causal contributions.
Outside the interaction, however, where values of $x$ or $x'$ less
than $\Lambda^{-1}$ have little support, the non-causal contributions
are suppressed.  For repulsive potentials ($g>0$), there is a
non-vanishing packet time advance, but its value is smaller than all
other scales in the problem.

\subsection{Effect of Pauli Principle}
Of course, the examples of two-particle scattering considered here
completely ignore the presence of the dense surrounding medium.  Apart
from an exact treatment of many-body correlation effects, an important
effect in a strongly interacting many-fermion system such as dense
nuclear matter comes from the fact that scattering into final states
below the Fermi surface is forbidden because the states are already
filled~\cite{Weisskopf}.  Thus, for an incident plane wave in one dimension 
with momentum below the Fermi
surface, forward scattering is permitted but backward scattering is
not.  In general, backscattered wave packets can only contain
components above the Fermi surface.  Such higher momentum components
will be less sensitive to the details of the nuclear two-particle
potential, and the 
link to non-causal effects will be commensurately reduced.

\subsection{Packet widths and observable non-causal behavior} While both
the local barrier and the Yamaguchi interaction violate a causal
condition via the spectral sum in Eq.~\ref{spectralsum}, only the local
barrier (finite or infinite) displays a backscattered wave packet
advance.  In the latter case, the consequences are in principle
observable, but one must construct a wave packet with properties
suitable for such an observation.  The distance scale of the violation
is the range $a$ of the interaction ({\it i.e.,} size of the barrier).
In order to observe such a violation for a single scattering, one would
need a packet whose overall width is comparable to $a$.  Thus, a
nucleon-nucleon core radius of 0.5~fm would correspond to a relevant
packet whose momentum spread is roughly 2~fm${}^{-1}=400$~MeV$/c$,
which is quite different from the characteristics of a particle
emerging from an accelerator.

To observe non-causal effects in a many-particle system, one could use
much wider packets, since the accumulated spacetime advance has a much
larger scale.  Such a packet would then have a very sharp momentum
distribution, and a qualitative picture emerges by considering
individual plane-wave states.  As noted above, only momentum components
above the Fermi surface contribute to backscattering.  Furthermore, if
the repulsive core is not singular, only a portion of the incident
packet will be backscattered.  This means that each successive
scattering will reduce the amplitude of the packet, making it much more
difficult to observe in a macroscopic system.

\section{Conclusions}

We have examined a number of issues associated with the nature of
causality in theories of dense matter.  A strong causality condition is
the vanishing of field commutators (or anticommutators for fermions) for
spacelike separations of spacetime.  This condition precludes any
observable non-causal behavior over arbitrarily small distance scales.
In field theories, this condition cannot be achieved without the
antiparticle contributions to the field.  It should however be noted
that the presence of antiparticles does not guarantee locality.  The
important observation is that the field itself is the appropriate local
observable (in the case of Bose   fields).   The particle excitations of
the field are not adequate for testing Einstein causality  
because there are problems in localizing particle
degrees of freedom within a Compton wavelength.  This problem, which
occurs on the same scale as the violations of local commutation
relations when antiparticles are dropped, is related to the inability to
define a position operator that transforms as a four-vector.  It is a
manifestation of the uncertainty principle.  In a sense,
a relativistic free particle behaves like a composite particle with its
size given by its Compton wavelength. 

If a dense medium is described via the direct interaction of particles
rather than fields, then there are apparent violations of causality on
the Compton wavelength scale.  This is not a real problem; it is related
to the inability to localize the particle within a Compton wavelength. 
However, without antiparticle degrees of freedom,  the theory has no
local operators that can be used to precisely test Einstein causality.  
More seriously, there are violations of causality on the scale of the
range interaction that are in principle observable.   These come from
true non-localities in the dynamics.  

Although our discussion and analysis was restricted to models of directly
interacting particles,  one can anticipate that similar violations of
causality can be identified in any truncation of local field theory
that leads to a violation of locality.  These include the use of
cutoffs, effective field approximations,  phenomenological quasiparticle
models, and Fock space truncations.  Clearly, one does not expect that
any experiment can lead to violations of Cauchy or Einstein causality;
however, our discussion suggests that for any given model, there
are in principle some limits to its applicability.  Empirically, we
find that typical models do not seem to lead to large violations of 
Einstein causality.

One of the questions which prompted this study is why conventional
nonrelativistic calculations of dense matter properties work as well as
they do with respect to causal requirements.  There appear to be
several reasons:
\begin{itemize}
\item coincidence: Equations of state $p(\epsilon)$ based upon
  nonrelativistic dynamics have no {\it a priori} reason either to
  satisfy or to violate a causality condition.  The dependence upon
  the speed of light of these calculations appears only trivially
  through the rest energy contribution to the energy density.
\item The densities of interest (say, even up to 10 times normal
  nuclear density) are still very low compared to the close-packed
  limit (closer to 100 times nuclear density) where any non-causal
  effects would be most pronounced.
\item Finite range and/or smooth potentials mitigate the time advance
  characteristic of backscattered wave packets.  Repeated
  backscattering greatly reduces a time-advanced packet of information over
  macroscopic distances.
\item Pauli principle: filled scattering states mean that the
  backscattered packet is either damped or eliminated.
\end{itemize}

\section{Acknowledgements} 
This work was prompted in part by a discussion between BDK and
Professor J. D. Walecka on this subject many years ago.  He also
thanks Professor B. D. Serot for a helpful conversation.  This work
was supported in part by 
the U.S. National Science Foundation under Grant PHY-9319641.
\pagebreak
\mediumtext

\pagebreak
\mediumtext
\begin{figure}
  \caption{Time advance of a two-particle wave packet in the
    center-of-mass frame.}  
\end{figure}

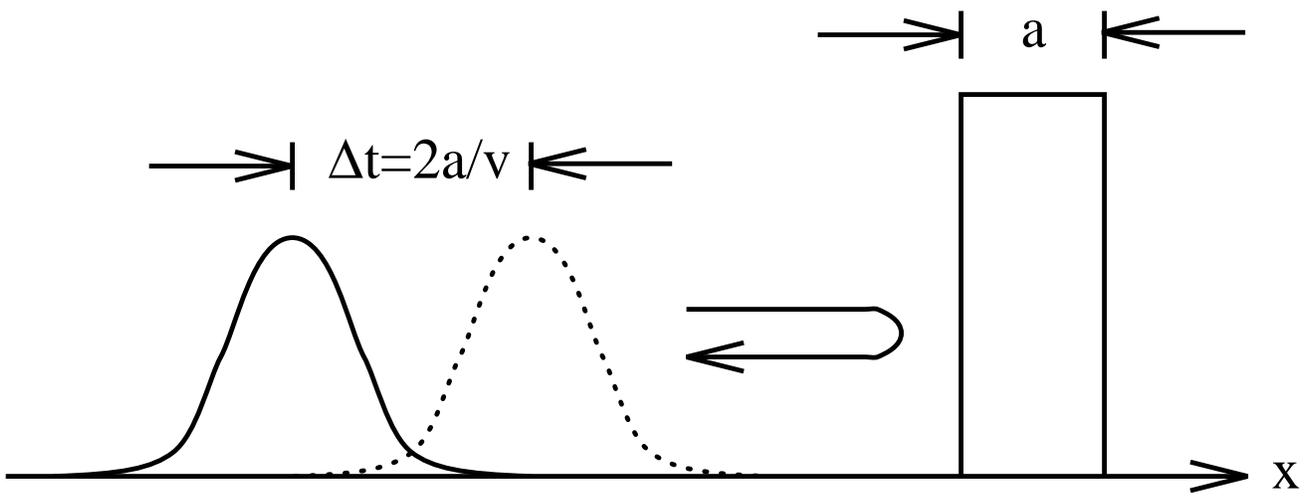
\end{document}